\documentstyle[12pt, fleqn]{article}
\begin{document}
\parskip=12pt
\baselineskip=20pt

\vspace{5cm}
{\begin{center}
{\Large{Integration of Einstein's equations in the weak field domain using the "Einstein" gauge}}  \\
\end{center}
\vspace{5cm}
\begin{center}
{{\sc E. Hitzer${}^{\dagger} $  and H. Dehnen${}^{\ddagger} $ } \\
{\it Physics Department\\
University of Konstanz \\
Box 5560\\
D-78434 Konstanz}}
\end{center}

\vspace{3cm}

\begin{center}
${}^{\ddagger} $ Email: hitzer@mech.mech.fukui-u.ac.jp\\
${}^{\dagger} $ Email: Heinz.Dehnen@uni-konstanz.de
\end{center}

\newpage

\section*{Abstract}

We propose a new alternative gauge for the Einstein equations instead of the 
de Donder gauge, which allows in the limit of weak fields a straightforward 
integration of these equations. The Newtonian potential plays a new 
interesting role in this framework. The calculations are demonstrated 
explicitely for 2 simple astrophysical models. 

\newpage 

\section*{I. Introduction}

The usual way to solve the Einstein equations after linearization
 is first to choose the de Donder- or harmonic gauge and second to
 integrate, in strict analogy to the inhomogeneous Maxwell
 equations of electrodynamics, via the use of retarded (in order
 to preserve causality) Green functions. 

Up to the present day the common opinion was that this would be
 the only general way to derive a comprehensive solution of the
 Einstein equations. This has led to many speculations about the
 uniqueness of the de Donder gauge, which many scientists tried to
 interpret not only as some random mathematical structure, but as
 to be more fundamental - the ``physical gauge''{\footnote{Compare Fock, 1964  \S \S \, 92, 93.}}. Further fundamental physical arguments for such an opinion
 have been given. 

Nevertheless in working with rather unconventional methods on
 finding the correct description for the microscopic gravitational interaction between elementary particles  in the realm of
 quantum physics (Dehnen and Hitzer, 1994 and 1995; Hitzer 1996; 
 Geitner, Hanauske and Hitzer, 1998) we were directed in a natural way to
 a new way of integrating the
 Einstein equations in their linearized version. This new method
 relies on basically two pillars: the use of a special gauge,
 which Einstein himself had suggested (Einstein, 1916) in order to give his theory an as
 elegant and concise shape as possible; and a supplementary gauge
 condition adding some further specification to Einstein's
 original suggestion. After implementing this supplemented special
 gauge, the integration of the Einstein equations after
 linearization becomes straightforward. So far as we know, such an integration is not discussed in the literature.

We have obtained an alternative to the widely accepted
 standard of the de Donder gauge, so that the arguments about its
 fundamental physical nature should be carefully reviewed, especially 
 since with the use of the new gauge Einstein's non-linear theory becomes 
 polynomial. This will be discussed in detail in a future paper.

Yet another highlight concerns the role of Newton's scalar
 potential which naturally appears in the Einstein equations in
 this gauge. No labourious and subtle procedure will have to be
 applied to regain Newton's scalar theory but its natural
 embedding easily unfolds. 

\section*{II. The Einstein equations in the supplemented Einstein gauge}

The general non Euclidean metric $g_ {\mu \nu } = \mbox {diag} (+, -, -, -) $ may be decomposed as: 
$$
g_ {\mu \nu } = \eta _ {\mu \nu } + \gamma _ {\mu \nu } , 
\eqno (2.1) 
$$
where $\eta _{\mu \nu }$ is the usual Minkowski metric (exclusively used for raising and lowering indices) and $| \gamma _ {\mu \nu }| << 1$ . In the Einstein-gauge we have
$$
\mbox {det} (g_ {\mu \nu } ) \stackrel{!}{=} -1 . 
\eqno (2.2)
$$
Inserting (2.1) into (2.2) yields (index raising and lowering only with $\eta _ {\mu \nu }$) 
$$
\gamma := \gamma _ {\mu \nu } \eta ^{\mu \nu } = 0. 
\eqno (2.3)
$$
Using (2.3), the Einstein equations 
linearized in $\gamma _ {\mu \nu }$ can be simplified to 
$$
\partial ^\alpha \partial _ \alpha \gamma ^{\mu \nu } - 
\partial _\alpha  \partial ^\mu  \gamma ^{\alpha \nu } - 
\partial _ \alpha \partial ^\nu  \gamma ^{\alpha \mu } + 
\partial _\alpha \partial _ \beta \gamma ^{\alpha \beta } 
\eta ^{\mu \nu } = 
- 16 \pi G T^{\mu \nu } . 
\eqno (2.4)
$$

We first show in which way the Einstein-gauge can be achieved initially assuming $\gamma \neq  0$. The general gauge transformations are
$$
x^\mu = x^{\prime \mu } + \xi ^\mu 
\eqno (2.5a)
$$
resulting infinitesimally in 
$$
\gamma ^{\prime} _ {\mu \nu } = 
\gamma  _ {\mu \nu } + 
\partial  _ \nu \xi _\mu + 
\partial _ \mu \xi _\nu , 
\eqno (2.5b)
$$
and 
$$
\gamma ^{\prime} = \gamma +  2 \partial  _\alpha \xi ^\alpha . 
\eqno (2.5c)
$$
Condition (2.3)  of the vanishing of $\gamma ^\prime $ may now be achieved  by demanding 
$$
\partial _\alpha \xi ^\alpha \stackrel{!}{=} 
- \frac{1}{2} \gamma . 
\eqno (2.6)
$$
But that is only one condition and we will prove right away that (2.6) may be supplemented by 
$$
\partial _\nu  \gamma ^{\prime} _ \mu  {}^\nu = 
\partial _\nu  \gamma _\mu  {}^\nu + 
\partial _\nu  \partial ^\nu  \xi _\mu + 
\partial _\mu  \partial _\nu  \xi ^\nu  
\stackrel{!}{=} 2 \partial _\mu f , 
\eqno (2.7)
$$
where $f$ is a scalar function of the coordinates. 
Inserting (2.6) into (2.7) yields:
$$
\partial _ \nu \partial ^\nu  \xi _ \mu  - 2 \partial _\mu  f 
= \frac{1}{2} \partial _\mu  \gamma  - \partial _\nu  \gamma  _\mu  {}^\nu .
\eqno (2.7a)
$$
Equations (2.6) and (2.7a) have the common feature, that the unknown entities $\xi ^\mu $ and $f$ are on the left-hand sides and the known ones on the right-hand sides. Their solution works as follows. Forming the divergence of (2.7a) yields
$$
\partial _\nu \partial ^\nu  \partial _\mu  \xi ^\mu  - 
2\partial _ \mu  \partial ^\mu  f = 
\frac{1}{2} \partial _\mu  \partial ^\mu  \gamma  - 
\partial _\mu  \partial _\nu  \gamma ^{\mu \nu } . 
\eqno (2.8)
$$
Inserting (2.6) a second time and rearranging the terms leads to the determination equation for $f$: 
$$
\partial _\mu  \partial ^\mu  f = 
\frac{1}{2}(\partial _\mu  \partial _\nu   \gamma^{\mu \nu }-  \partial _\mu  \partial ^\mu  \gamma  ). 
\eqno (2.9)
$$
Equation (2.9) allows us immediately to calculate $f$ via Green functions. The solution for $f$ may now be inserted into (2.7a) and the vector $\xi ^\mu $ therefore can be explicitely calculated from 
$$
\partial _\nu  \partial ^\nu  \xi _ \mu = 
2\partial _\mu  f + 
\frac{1}{2} \partial _\mu  \gamma  - \partial _\nu  \gamma _\mu  {}^\nu 
\eqno (2.10)
$$
via Green functions. The solutions for $f$ and $\xi ^\mu $ will obviously satisfy (2.6) and (2.7) and therefore allow us to work in a coordinate system in which in addition to the Einstein-gauge (2.2) or (2.3) equation  (2.7) holds. Having achieved this, we may rewrite the Einstein equations (2.4) using (2.3) and (2.7) as 
$$
\partial _\alpha \partial ^\alpha \gamma ^{\mu \nu } 
- 4 \partial ^\mu  \partial ^\nu  f + 2 
\partial _\alpha  \partial ^\alpha  f \eta ^{\mu \nu } = 
- 16 \pi  G T ^{\mu \nu } 
\eqno (2.11a)
$$
and the trace equation
$$
\partial _\alpha  \partial ^\alpha  f = 
- 4 \pi G T 
\eqno (2.11b) 
$$
($T$ being the trace of $T^{\mu \nu } $). Because of $\partial _\nu  T_ \mu  {}^\nu = 0$ the conditions (2.3) and (2.7) are in turn consequences of the field equations (2.11) as it is the case in the de Donder-gauge. Thus the conservation law guarantee the existence of the retarded or advanced integrals of (2.9) and (2.10).  

\section*{III. The new integration procedure }

The general way to solve Einstein's equations (2.11) bearing the 
special gauge (2.3) in mind will be first to calculate the scalar 
function $f$ via convolution of the trace of the energy momentum 
tensor on the right-hand side of (2.11b) with the well known retarded Green functions $D(x - x^\prime)$ of the D'Alembert operator on the left 
hand side: 
$$
f(x) = 4 \pi G \int D(x - x^\prime) T(x^\prime) d^4 x^\prime . 
\eqno (3.1)
$$
The solution for $f$ must than be inserted on the left of (2.11a) 
in order to obtain: 
$$
\partial _ \alpha \partial ^\alpha \gamma ^{\mu \nu } = 
 - 16 \pi G \left[ T^{\mu \nu } - \frac{1}{2} \eta ^{\mu \nu } T 
- \partial ^\mu  \partial ^\nu   \int D (x - x^\prime ) 
T( x^\prime ) d^4 x^\prime \right]. 
\eqno (3.2)
$$
A second integration via convolution of the right-hand side of (3.2) 
with the respective Green functions of the D'Alembert operator on the 
left-hand side results in this final explicit expression for the non-Euclidean 
deviation $\gamma ^{\mu \nu }$ of the general metric from the 
Minkowski metric: 
$$
\gamma ^{\mu \nu } =  16 \pi G 
\int D(x - x^\prime ) 
\Bigg\{  T^{\mu \nu } (x ^\prime ) - 
$$
$$
- 
 \frac{1}{2} \eta ^{\mu \nu } T( x^\prime ) - \partial ^{\mu \prime} \partial ^{\nu \prime} \int D( x^\prime - x^{\prime \prime} ) 
T( x^{\prime \prime} ) d ^4x^{\prime \prime} \Bigg\} d^4  x^\prime . 
\eqno (3.3)
$$
It is remarkable, that in the vacuum $(T_ {\mu \nu } = 0)$ equ. (3.2) does not go over into a D'Alembert equation, as it is the case in the de Donder-gauge. Consequently the general conditions on $T_ {\mu \nu }$ for the existence of the integral in (3.3) cannot be given so easily. Instead of this we discuss in chapter V two examples.

\section*{IV. The Newtonian limit}

In the Newtonian limit we simply approximate the D'Alembert operator by the 3-dimensional space Laplace operator and the trace $T$ of the energy momentum tensor on the right-hand side of (2.11b) by the matter density $\rho $: 
$$
\Delta  f = 4 \pi G \rho , 
\eqno (4.1)
$$
which reveals that the scalar function $f$ must be identified with  Newton's scalar potential $\phi$: 
$$
f(x) =  \phi (x) . 
\eqno (4.2)
$$
It is therefore obvious that in the presently proposed specified ``Einstein'' gauge the trace of the Einstein equations (2.11b) is the general relativistic analogue of the scalar Newtonian equation of gravity. It is shown thereby that a ``generalized form'' of Newton's scalar potential, the scalar $f$, is naturally present in the theory of general relativity. We now rewrite the time-time component of equation (3.2) for the static case: 
$$
 \Delta \gamma ^{00}=  
16 \pi G ( T^{00} - \frac{1}{2} T) . 
\eqno (4.3)
$$
Replacing $T^{00}$ by the matter density $\rho $ yields the conclusion, that   $\gamma ^{00}$ is two times $f$ and therefore:
$$
\gamma ^{00} (x) = 2f(x) = 2 \phi (x) 
\eqno (4.4)
$$
Now, according to the geodesic equation this $\gamma ^{00}$ suffices to determine the nonrelativistic trajectory of a massive body. 

Regarding the swiftness and elegance of this transition from
 General Relativity to Newtonian gravity it seems to work 
 just as natural as in the frame work of the de Donder 
 gauge. Newton's scalar potential itself aspires to new, 
 not only non relativistic eminence. 

\section*{V. Solutions for homogeneous and polytropic spheres}

Since gas spheres with homogeneous density $\rho $ may serve as simple models 
 for stars or other astrophysical objects, the
 solution of Einstein's equations in the presently proposed
 ``Einstein'' gauge will be given explicitely. A second more
 realistic model, the polytropic gas sphere with zero pressure on its surface,
 follows. 

\subsection*{5.1 Gas spheres with homogeneous density $\rho $} 
In the case of static homogeneous density 
$$
\rho, (\, \vec x) = 
\left\{  
\begin{array}{ll}
\rho, & | \, \vec x| \leq R \\[1ex]
0, & |\, \vec x| > R 
\end{array}
\right.
\eqno (5.1)
$$
in first approximation the energy-momentum Tensor $T^{\alpha \beta } $ takes the simple form 
$$
T^{\alpha \beta } = 
\left\{  
\begin{array}{lll}
\rho  & \mbox {for} & \alpha = \beta  = 0 \\[1ex]
0   & \mbox {for} & \alpha , \beta  \neq 0 
\end{array}
\right. 
\eqno (5.2)
$$
with the trace 
$$
T = T^{\alpha \beta } \eta _ {\alpha \beta } = \rho . 
\eqno (5.3)
$$
According to (4.2) and (4.4), $f$ becomes the usual Newtonian scalar potential and $\gamma ^{00} $ its double value,
$$
\gamma _ {00} (x) = 2 f(x) = 
\left\{  
\begin{array}{ll}
\frac{M G}{R} ( \frac{r^2}{R^2} - 3 ) & r < R \\[1ex]
- 2 \frac{M G}{r}  & r > R 
\end{array}
\right. 
\eqno (5.4a)
$$
where 
$$
M = \frac{4}{3} \pi R^3 \rho  ; \quad r = |\, \vec x| . 
\eqno (5.5)
$$
The space-time components of $\gamma ^{\alpha \beta } $ are zero, 
$$
\gamma _ {0 \mu } = \gamma _ {\mu 0} = 0 \quad \mbox {for} \quad \mu  = 1,2,3 .
\eqno (5.4b)
$$
The space components are, according to (3.3),
$$
\gamma _ {\mu \nu } (x) = 
\delta _ {\mu \nu  } \left[\frac{1}{3} f(x) + 
h (x) \right] - 3 \frac{x ^ \mu x ^ \nu }{r^2} h (x), 
\quad 
\mu , \nu \in {1,2,3}
\eqno (5.4c)
$$
where 
$$
h(x) = 
\left\{  
\begin{array}{ll}
\frac{4}{15} MG \frac{r^2}{R^3} & r < R \\[1ex]
\frac{2}{3} MG \frac{1}{r} - \frac{2}{5} MG \frac{R^2}{r^3} 
&  r > R  . 
\end{array}
\right. 
\eqno (5.4d)
$$
It is interesting that for $R > 0 $ the gravitational potentials (5.4c) contain for $r > R $ not only $1/r $-terms but also such terms which decrease with $1/r^3 $. Equation (5.4a) already reveals, that the gravitational redshift turns out as usual. 

The solution for $r > R $ agrees with Einstein's original suggestion for the field of a point mass (Einstein, 1916, p. 819){\footnote{Einstein omitted the factor 2 in his formulas.}} 
 
$$
\gamma _ {\mu \nu Einstein} = 
\left\{  
\begin{array}{ll}
- 2 \frac{GM}{r}, & \mu  = \nu  = 0 \\[1ex]
- \frac{2 G M }{r} \frac{x ^\mu  x^\nu }{r^2}, & \mu ,\nu , \epsilon \left\{  1,2,3 \right\} \\[1ex]
0, & \mbox {otherwise} 
\end{array}
\right. 
\eqno (5.6)
$$
up to a term
$$
a_ {\mu \nu } = 
\left\{  
\begin{array}{ll} 
\frac{2}{5} G M R^2 \left( - \delta _ {\mu \nu } \frac{1}{r^3} + 
3 \frac{x ^\mu  x ^\nu }{r^5} \right)   & \mu ,\nu \in \left\{  1,2,3 \right\} \\[1ex]
0 & \mbox {otherwise} 
\end{array}
\right. 
\eqno (5.7)
$$
which is such that 
$$
\Delta a_ {\mu \nu } = 0
\eqno (5.8)
$$
and which vanishes in the point mass limit $R = 0$. Following the original procedure Einstein  used to calculate the gravitational bending of light via the Huygen's principle (Einstein, 1916) {\footnote{Because Einstein missed a factor 2 in the components of his metric [Einstein, (1916), formula (70)], he also 
obtained only one-half of the correct value for the light bending angle 
(Bergmann, 1942, p. 221).}} 
applied to the metric (5.6) and (5.7) yields the usual result. 

\subsection*{5.2 The polytropic gas sphere}

The general polytropic equation of state  is:
$$
p = \alpha \rho ^\gamma ; \quad \alpha , \gamma  = \mbox{const}. \quad \alpha > 0, \gamma \geq 1 
\eqno (5.9)
$$
where p and $\rho $ are pressure and density of the gas, respectively, and $\gamma $ is the so called polytropic index. 
It is known that in the Newtonian case the Emden equation is exactly solvable for the physically interesting value $\gamma  = 2$ (e.g. matter inside Jupiter). Therefore we restrict ourselves in the following to that case $\gamma = 2.$  Then 
it can be shown (Dehnen and Obregon, 1971) that the conservation laws yield exactly 
$$
\rho = \frac{1}{\alpha } \left[ \left(   
\frac{\xi  _s }{\xi } \right) ^\frac{1}{2} - 1 \right] 
\eqno (5.10)
$$
where $\xi $ is the length of the timelike Killing vector{\footnote{$|| \nu $ means the covariant differentiation w.r.t. $x^\nu $.}}:
$$
\begin{array}{c}
\xi _ {\mu || \nu } + \xi _ {\nu || \mu } = 0 \\[1ex]
\xi ^2 := \xi _ \mu  \xi ^\mu  > 0 .
\end{array}
\eqno (5.11)
$$
The index $s$ in (5.10) indicates the evaluation at the surface of the 
sphere $r = R$. 

The coordinates  can be chosen such that 
$$
\xi = \sqrt {g_ {00}} = \sqrt {1 + \gamma _ {00}} 
\eqno (5.12) 
$$
which yields for $\rho $ the approximation $(| \gamma _ {\alpha \beta }| << 1)$:
$$
\rho = \frac{1}{4 \alpha }  (\gamma _ {00s} - \gamma _{00} ) . 
\eqno (5.13)
$$
The equation of state (5.9) shows for $\gamma =2$ that the pressure vanishes in first order approximation. We again have the same structure of $T^{\alpha \beta }$ as we had in (5.2) with the only difference that $\rho $ itself does now depend on the metric. 

The solution now is 
$$
 \gamma _ {00} (x) = 2 f(x) = 
\left\{  
\begin{array}{ll} 
- 2 \frac{MG}{R} (1 + \frac{R}{\pi r} \sin \frac{\pi r}{R}), 
& r < R \\[1ex]
- 2 \frac{MG}{r} & r > R . 
\end{array}
\right. 
\eqno (5.14a) 
$$
(5.4b) is again valid for the space time components. 

The space-space components are again given by (5.4c).  Only the function $h$ is now:
$$
h(x) = 
\left\{  
\begin{array}{ll} 
- 4 \frac{MG}{R} \left\{   
- (\frac{R}{r \pi })^3 
\sin ( \frac{\pi r}{R} ) + 
(\frac{R}{r \pi } )^2 
\cos (\frac{\pi r}{R }) + 
\frac{2}{3} (\frac{R}{r \pi }) \sin 
(\frac{\pi r}{R}) \right\},  & r < R \\[1ex]
- 2 \frac{M G}{r} \left\{  - 2 (\frac{R}{r \pi } )^2 + 
\frac{1}{3} \frac{R^2}{r^2} - \frac{1}{3} \right\},  & r > R . 
\end{array}
\right. 
\eqno (5.14b) 
$$
The solution for $r > R $ again agrees with Einstein's suggestion (5.6) up to a term $a_ {\mu \nu }$ which is identical with (5.7) up to a simple change in the numerical factor: 
$$
\frac{2}{5} \rightarrow 2 (\frac{1}{3} - \frac{2}{\pi ^2} ) 
\eqno (5.15)
$$
The remarks below equations (5.4), resp. (5.8), also apply for the gravitational metric of the polytropic gas sphere calculated above. It may be of interest that for other (finite) values of the polytropic index no analytic solution can be given. 

\subsection*{5.3 Equivalence with the Schwarzschild metric} 

In order to prove in both cases the equivalence of the calculated metrics for $r > R$ outside the spheres with the Schwarzschild metric (radial coordinate $r_S $) one needs to write both metrics in polar coordinates. Rescaling the radial coordinate by 
$$
r ^2 _S = r^2  (1 + F G M \frac{R^2}{r^3 } ) 
\eqno (5.16a)
$$
with
$$
F = 
\left\{  
\begin{array}{cc}
\frac{2}{5} & \mbox{ for the homogeneous gas sphere} \\[1ex]
2 (\frac{1}{3} - \frac{2}{\pi } ) & \mbox{for the  polytropic  gas sphere}
\end{array}
\right.
\eqno (5.16b)
$$
yields the usual form of the Schwarzschild metric in 
first-order approximation. 
\section*{VI. Conclusions}

The proposed new gauge and the new way for solving the Einstein equations may seem in the weak field domain only as some kind of technical alternative apart from the new light they shed on the role played by the Newtonian potential even in the theory of general relativity. But the general nonlinear  form of the special ``Einstein'' gauge is under investigation. The results will be presented in a future paper. 

\section*{References}

\begin{itemize}
\item[] Bergmann, P.G., (1942). Introduction to the Theory of Relativity, Prentice-Hall, Inc. Englewood Cliffs, N.J.

\item[] Dehnen, H. and Obregon, O. (1971). Astronomy and Astrophysics \underline{12}, 161. 

\item[] Dehnen, H. and Hitzer, E. (1994). International Journal of theoretical Physics \underline{29}, 537.

\item[] Dehnen, H. and Hitzer, E. (1995). International Journal of theoretical Physics, \underline{34}, 1981.

\item[] Einstein, A. (1916). Annalen der Physik \underline{49}, 769. 

\item[] Fock, V. (1964). The Theory of Space, Time and Gravitation, Pergamon Press, Edinburgh. 

\item[] Geitner, A., Hanauske, M. and Hitzer, E. (1998). Acta Physica Polonica, \underline{B29}, 971.

\item[] Hitzer, E. (1996). The Higgs-Field Theoretic Extension of the Spin-Gauge Theory of Gravity, Series: Konstanzer Dissertationen, \underline{501}, Hartung-Gorre, Konstanz.

\end{itemize} 

\end{document}